\documentclass{elsart3p}
\usepackage{graphicx}
\usepackage{amsmath}
\usepackage{amssymb}

\begin{document}

\begin{frontmatter}

\title{Designed defects in 2D antidot lattices for quantum information processing}

\author[address1]{Jesper Pedersen\thanksref{thank1}},
\author[address1]{Christian Flindt},
\author[address1]{Niels Asger Mortensen},
and
\author[address1,address2]{Antti-Pekka Jauho}

\address[address1]{MIC -- Department of Micro and Nanotechnology,
             NanoDTU, Technical University of Denmark, Building 345east,
             DK-2800 Kongens Lyngby, Denmark}
\address[address2]{Laboratory of Physics, Helsinki University of Technology, P. O. Box 1100, FI-02015 HUT, Finland}

\thanks[thank1]{
Corresponding author.\\
E-mail: jesper.pedersen@mic.dtu.dk}

\begin{abstract}
We propose a new physical implementation of spin qubits for quantum information processing, namely defect
states in antidot lattices defined in the two-dimensional electron gas at a semiconductor heterostructure.
Calculations of the band structure of a periodic antidot lattice are presented. A point defect is created
by removing a single antidot, and calculations show that localized states form within the defect, with
an energy structure which is robust against thermal dephasing. The exchange
coupling between two electrons residing in two tunnel-coupled defect states is calculated numerically.
We find results reminiscent of double quantum dot structures, indicating that the suggested structure is
a feasible physical implementation of spin qubits.
\end{abstract}

\begin{keyword}
EP2DS-17 \sep quantum computing\sep exchange coupling \sep antidot lattices
\PACS 03.67.Lx \sep 73.21.Cd \sep 75.30.Et
\end{keyword}
\end{frontmatter}

The possibility of utilizing the spins of electrons confined in quantum dot systems
as the fundamental building blocks for large-scale quantum computing was first introduced
by Loss and DiVincenzo in 1998~\cite{Loss:1998} and has since led to numerous theoretical
and experimental studies within this field~\cite{Burkard:1999,Gorman:2005,Helle:2005,Koppens:2005,Petta:2005,Zhang:2006}.
In the proposal by Loss and DiVincenzo the exchange coupling between the spins of the electrons serves as the
mechanism for coherent manipulation of and interaction between the spin qubits. Inspired by
these ideas we have recently proposed to use bound states which form at the location of
point defects in periodic antidot lattices as an alternative way of realizing spin
qubits~\cite{Flindt2005}. The available fabrication methods suggest that such structures
may offer high scalability more readily than conventional gate-defined quantum dots~\cite{Fuhrer2002}.

We consider a two-dimensional electron gas superimposed with a triangular lattice of antidots
with lattice constant $\Lambda$. In the effective-mass approximation the two-dimensional
single-electron Hamiltonian is
\begin{equation}
H = -\frac{\hbar^2}{2m^*}\nabla_\mathbf{r}^2+\sum_i V\left(\mathbf{r}-\mathbf{R}_i\right),\;
\mathbf{r}=(x,y),
\end{equation}
where $m^*$ is the effective mass of the electron and $V(\mathbf{r}-\mathbf{R}_i)$ is the
potential of the $i$'th antidot positioned at $\mathbf{R}_i$. We use parameter values typical
of GaAs, for which $\hbar^2/2m^*\simeq 0.6$ eVnm$^2$, and assume a lattice constant of
$\Lambda = 45$ nm. Modeling each antidot as an infinite circular potential barrier of diameter $d$
allows us to solve the problem using finite-element methods with the Dirichlet boundary condition
that the eigenfunctions are zero in the antidots~\cite{Flindt2005}. 

\begin{figure}[tbp]
\begin{center}
\includegraphics[width=.9\linewidth]{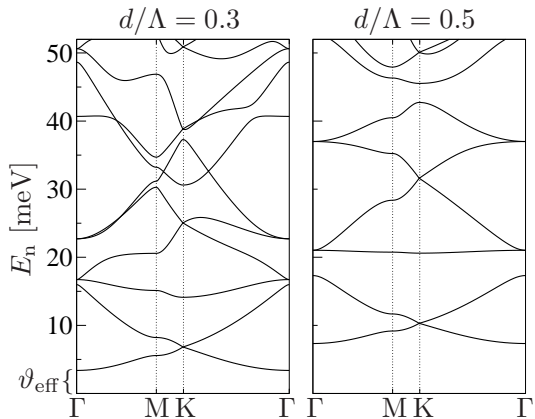}
\caption{Band structure of the periodic antidot lattice with lattice constant $\Lambda = 45$ nm and
two different values of the relative antidot diameter $d/\Lambda$. On the left graph the
gap $\vartheta_\mathrm{eff}$ is indicated, below which no states exist for the periodic structure.} \label{fig:band}
\end{center}
\end{figure}
The calculated band structure of the periodic antidot lattice is shown in Fig.\ \ref{fig:band} for two
different values of the relative antidot diameter $d/\Lambda$. Also indicated on the figure is
the gap $\vartheta_\mathrm{eff}$ below which no states exist for the periodic structure. 
The band gap around 20 meV is present for $d/\Lambda > 0.35$ while the higher-energy band gap
around 45 meV only exists for $d/\Lambda > 0.45$. The existence and location of all band gaps have
been verified by density of states calculations (not shown)~\cite{Pedersen2:2007}.
The general increase in energies with the antidot diameter is due to the increased confinement of the Bloch states.

\begin{figure}[tbp]
\begin{center}
\includegraphics[width=\linewidth]{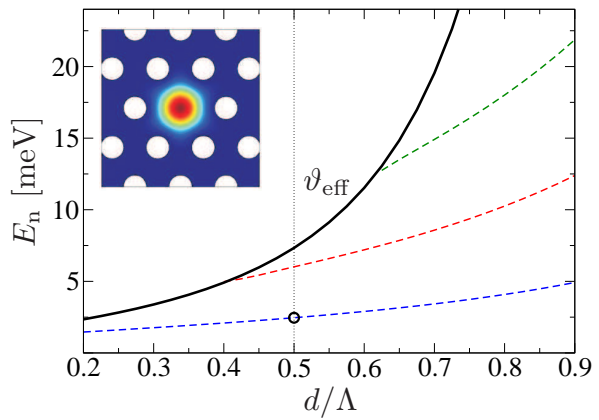}
\caption{Energy spectrum for a single defect, showing the three lowest energy eigenvalues
as a function of the relative antidot diameter $d/\Lambda$. The full line indicates the height
$\vartheta_\mathrm{eff}$ of the effective potential in which the localized states reside. The
inset shows the absolute square of the localized eigenfunction corresponding to the eigenvalue
indicated with a circle.}
\label{fig:defects}
\end{center}
\end{figure}

We next consider the case where a point defect has been introduced in the lattice by leaving out a single
antidot. The gap $\vartheta_\mathrm{eff}$ defined in Fig.\ \ref{fig:band} may be considered as the
height of an effective two-dimensional circular  step potential surrounding the defect, and thus gives
an upper limit to the existence of bound states localized in the defect. Similar states are expected
to form in the band gaps of the periodic structure. We refer to these localized states as defect states.
These decay to zero far from the location of the defect, allowing us to solve the problem on a
domain of finite size, imposing once again Dirichlet boundary conditions on the antidots and on the
edge of the domain. 
The discrete spectrum of a single defect is shown in Fig.\ \ref{fig:defects} for states residing
below $\vartheta_\mathrm{eff}$. The inset shows the eigenfunction corresponding to the lowest
eigenvalue. As expected a defect leads to the formation of a number of localized states at the location
of the missing antidot. Calculations have confirmed the existence of similar states in the band gap regions \cite{Pedersen2:2007}.
The results indicate that the number of localized states can be tuned via the relative antidot diameter
$d/\Lambda$, allowing for $n=1,2,3\dots$ levels in the defect. For $d/\Lambda = 0.5$ the energy
splitting between the two lowest states is approximately 3.6 meV, which is much larger than $k_BT$ at
subkelvin temperatures, and the energy structure is thus robust against thermal dephasing.

Together with single-qubit operations, the exchange coupling between the spins of electrons confined
in double quantum dot structures has been shown to be a sufficient mechanism for implementing a
universal set of quantum gates for quantum information processing~\cite{DiVincenzo2000}.
The exchange coupling arises as a consequence of the Pauli principle, which couples the symmetries
of the orbital and spin degrees of freedom. The splitting of the lowest eigenvalue $E_S$ corresponding to
a symmetric orbital wavefunction and the lowest eigenvalue $E_A$ corresponding to an anti-symmetric
orbital wavefunction may thereby be mapped onto an effective Heisenberg spin Hamiltonian
$\mathcal{H}=J\mathbf{S}_1\cdot\mathbf{S}_2$, where $J=E_A-E_S$ is the exchange coupling.

\begin{figure}[tbp]
\begin{center}
\includegraphics[width=\linewidth]{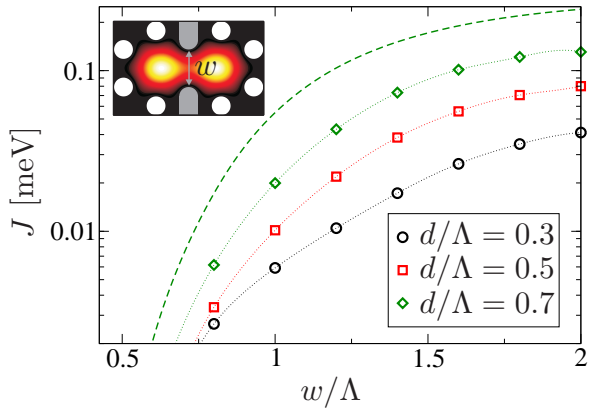}
\caption{Exchange coupling $J$ as a function of the relative split gate constriction
width $w/\Lambda$ for three different values of the relative antidot diameter $d/\Lambda$.
The dashed line indicate the results obtained in the Hubbard approximation for $d/\Lambda=0.7$.
The inset shows the calculated two-electron charge density of the singlet ground state
for $d/\Lambda=0.5$ and $w/\Lambda=2$. The split gate is shown in grey, while antidots
are white.}
\label{fig:Jvsw}
\end{center}
\end{figure}
Analogous to a double quantum dot system we now consider an antidot lattice in which a single antidot
and one of its next-nearest neighbors have been left out of the lattice. In the following we tune the
coupling between the defects via a metallic split gate defined on top of the 2DEG in order to control
the opening between the two defects. By increasing the applied voltage one squeezes the opening. As the exchange
coupling depends on the overlap of the defect states, we may thereby control the exchange coupling
electrostatically. The split gate is modeled as an infinite potential barrier shaped as
shown in the inset of Fig.~\ref{fig:Jvsw}.

Using recently developed numerically exact methods~\cite{Pedersen2:2007}, we have calculated the exchange coupling for
such a double defect geometry. In Fig. \ref{fig:Jvsw} we show the calculated exchange coupling as a
function of the relative split gate constriction width $w/\Lambda$ for three different values of the
antidot diameter. We also show the calculated two-electron charge density of the singlet ground state for
$d/\Lambda=0.5$ and $w/\Lambda=2$. The exchange coupling varies several orders of magnitude as the
split gate constriction width is increased. These results are similar to those obtained for double
quantum dot structures where the exchange coupling has been calculated as a function of
interdot distance~\cite{Burkard:1999}. The figure also shows the exchange coupling calculated in the Hubbard approximation,
$J_\mathrm{H} = 4t^2/U$, where $t$ is the tunnel coupling between the defect states while $U$
is the on-site Coulomb repulsion~\cite{Burkard:1999}. We note that while this approximation yields
qualitatively correct results for the entire range of parameters, the approximation clearly has no
quantitative predictive power. We also note that in general the validity
of the approximative schemes used to evaluate the exchange coupling in low-dimensional nanostructures
is highly dependent on both the form of the potential under consideration as well as the
choice of parameter values~\cite{Pedersen2007}.

In conclusion, we have shown that defect states in antidot lattices may serve as a physical
implementation of spin qubits for large-scale quantum information processing. We find that
introducing a point defect in an antidot lattice leads to the formation of localized states
within the defect, with a level structure which is robust against thermal dephasing.
Calculations of the exchange coupling show results similar to those obtained for double
quantum dot structures, allowing for electrostatic tuning of the exchange coupling over several orders of
magnitude.

\bibliographystyle{elsart-num}

\end{document}